\documentclass[11pt]{article}
\usepackage[utf8]{inputenc}
\usepackage{amsmath}
\usepackage{amssymb}
\usepackage{psfrag}
\usepackage{bbold}
\usepackage{graphicx}
\usepackage[text={6.5in,9in},centering]{geometry}
\usepackage{siunitx}
\usepackage{color,soul}
\usepackage{comment}
\usepackage{nicefrac}
\graphicspath{{./Figures/}}
\usepackage{subfigure}
\usepackage{graphicx} 
\usepackage{float} 
\usepackage{caption} 
\usepackage{tabularx} 
\usepackage{geometry} 
\usepackage{amsmath} 
\usepackage{breqn}
\usepackage{amssymb}
\usepackage{mathbbol}
\usepackage{textgreek}
\usepackage{titlesec} 
\usepackage{parskip} 
\usepackage{multirow} 
\usepackage{multicol}

\usepackage[T1]{fontenc} 

\title{Multimodal Resonance in Strongly Coupled Inductor Arrays}

\author{Robert R. Hughes, James Treisman, Alexis Hernandez Arroyo, Anthony J. Mulholland}

\date{DRAFT \today}

\begin{document}

\maketitle

\begin{abstract}
Magnetic resonance coupling (MRC) is widely used for wireless power transfer (WPT) applications, but little work has explored how MRC phenomena could be exploited for sensing applications.  This paper introduces, validates and evaluates the unique multi-resonant phenomena predicted by circuit theory for over-coupled inductive arrays, and presents eigen-formulae for calculating resonant frequencies and voltage modes within passively excited arrays. Finite-element simulations and experimental results demonstrate the validity of the multi-modal resonant principles for strongly-coupled inductor arrays.
The results confirm the distinctive multi-modal resonant frequencies these arrays exhibit, corresponding to the specific magnetic excitation "modes" (comparable to vibrational modes in multi-degree-of-freedom systems).
The theoretical and finite element models presented offer a framework for designing and optimizing novel inductive sensing arrays, capitalizing on the unique resonant effects of over-coupling and exploiting their potential magnetic field shaping. 

\end{abstract}

\section{Introduction}
Electrical resonance is a fundamental effect used in electrical engineering, physics and electronics that has been widely researched and utilized for many applications. 
The phenomenon refers to the ability of an electrical system, typically containing components that store and discharge energy between electric (capacitors) and magnetic (inductors) fields, to respond to an alternating voltage or current, and exhibit resonant behavior at a specific frequency \cite{Griffiths2008}. 
Over the past century, electrical resonance has been used in a wide variety of applications, ranging from simple devices such as radio and television tuners \cite{Blanchard1941}, to near-field sensing applications \cite{owston1970high, Hughes2014}, as well as more advanced applications such as wireless power transfer \cite{kurs2007wireless}.

It is in wireless power transfer (WPT) systems in particular that the concepts of multi-coil magnetic resonance coupling (MRC) systems have been explored predominatly for the transmission of power over greater distances \cite{Zhong2013} and between multiple inductive elements \cite{Liao2019, Hou2021, Liu2024}. It is in such systems that resonance splitting phenomena have been most commonly observed and exploited. In this paper, these same frequency splitting effects are considered for parallel LC measurements of coplanar inductor arrays, specifically evaluating the phenomenona in mm-scale inductive array sensors, comparable to sensor arrays used in non-destructive testing (NDT) applications.



When it comes to inductive sensor arrays, such as those used in non-destructive testing (NDT) applications, resonance is strictly avoided. This is due to the electronically unstable nature of resonance, while expensive electronic multiplexing is employed in order to avoid the unpredictable mutual interactions (cross-talk) between neighbouring coils. While the stability of avoiding resonance is beneficial, the potential advantages and novel technology enabled by understanding the behaviour of resonance in arrays sensors are ignored. 

Over the past decade, the resonance principles of WPT in multi-coil sensors has gained increasing interest as a potential area for novel technology development in near-field sensing applications \cite{hughes2016, Babu2016, Daura2023, song2021wireless}. Researchers have exploited the resonance splitting phenomenon to evaluate displacement between sensing coils \cite{Babu2018, arroyo2023displacement}, and even explored the effects of series resonant array sensors for non-destructive applications \cite{Daura2020}. However, while these effects have been recognised and exploited in series resonant arrays, a detailed evaluation of the phenomena experienced by arrayed resonant inductors is yet to be available from the literature for measurements for parallel LC resonators. This limits the potential for resonant multi-coil sensor design and optimisation to advance the field of resonant inductor array sensing.

In this paper, we present the theory, experimental validation and key observations behind the unique phenomena found in magnetically over-coupled inductive arrays exhibiting multiple resonant frequencies (multi-modal) for parallel resonant driver coils in passive LC resonant arrays. 
The paper covers both the theoretical principles of multi-modal electrical resonance arrays, comparison against finite element simulations and validation of experimental results.  While the study limits itself to the evaluation of co-planar array elements, the principles presented would naturally extend into two and three-dimensional array configurations.



\section{Theory}\label{sec:theory}
Electrical resonance occurs when an electrical system oscillates at its resonant frequency, $f_0$, as determined by its electrical properties of capacitance, $C$, inductance, $L$, and resistance, $R$. The classical expression for the natural resonant frequency of a electrical resonator with either series or parallel $L$ and $C$ components is defined as,
\begin{equation}
    \omega_0 = 2\pi f_0 \approx \sqrt{\frac{1}{LC}}, \label{eqn:w0}
\end{equation}
where $\omega_0$ is the natural resonant angular frequency of the system \cite{Griffiths2008}. The principles of exploiting the resonant frequency shifting behaviour of single inductive sensors has been explored by many authors \cite{owston1970high, Hughes2018}, however little has been done to explore the resonance phenomena of multi-sensor arrays.  

The influence of neighbouring inductor resonators on the electrical impedance spectra of a driven coil has been highlighted as a point of interest for applications in array sensing \cite{hughes2016, Daura2020}. 
The following sections detail the theory behind the resonance phenomena of magnetically-coupled multi-coil sensors.

\subsection{Two Coil Model}\label{sec:2coil}
When close to another comparable coil, a primary coil will inductively couple to a secondary coil. The effect of the secondary coil can be modelled as an inductor, $L_2$, in series with a resistor, $R_2$, and a capacitor, $C_2$.  This coupling, parameterised by the coupling coefficient, $k$, will alter the effective inductance and resistance ($L_1'$ and $R_1'$ respectively) of the primary measurement circuit and will distort the impedance. 
The full formulae for the impedance of the bimodal coil system can be found in \cite{arroyo2023displacement,hughes2015thesis}. 
The general expression for the resonant frequencies of a two-coil system can be defined as \cite{arroyo2023displacement},

\begin{equation}
\omega_{\pm}'{}^2 = \frac{1}{2}\frac{\omega_1^2}{\left[1-k^2\right]}\left[ 1 + \frac{\omega_2^2}{\omega_1^2}\left(1 \pm \frac{\omega_1}{\omega_2} \sqrt{\frac{\omega_2^2}{\omega_1^2} + \frac{\omega_1^2}{\omega_2^2} + 4k^2 - 2} \right) \right], \label{eqn:f0m1}
\end{equation}

where $\omega_n = \sqrt{1/L_n C_n}$.  Equation~\ref{eqn:f0m1} therefore predicts two unique solutions to the resonant frequency of the coupled system. Therefore, by knowing the natural resonant frequency of any two coils in their uncoupled ($k=0$) state, values for these two unique resonant frequencies can be determined for a given value of $k$. Equation~\ref{eqn:f0m1} can also be written in its dimensionless form for the frequency ratio relative to the resonant frequency of the driver coil $r_1 = \omega / \omega_1$,
\begin{equation}
r_{1}^2 = \frac{\left[ 1 + \frac{\omega_2^2}{\omega_1^2} \right] \pm \sqrt{\left[ 1 + \frac{\omega_2^2}{\omega_1^2} \right]^2 - 4(1-k^2) } }{2\left[ 1-k^2 \right]}. \label{eqn:f0m2}
\end{equation}


\begin{figure}[!b]
\centering
\includegraphics[width=3.0in]{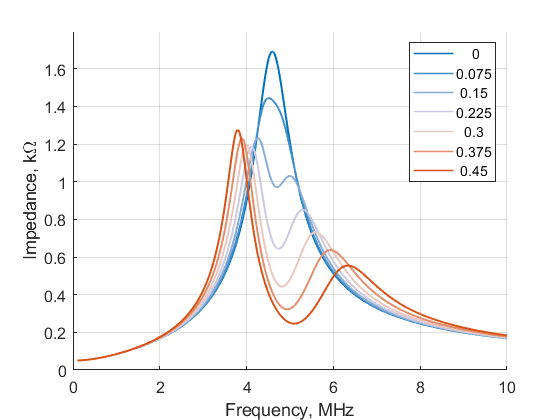}
\includegraphics[width=2.5in]{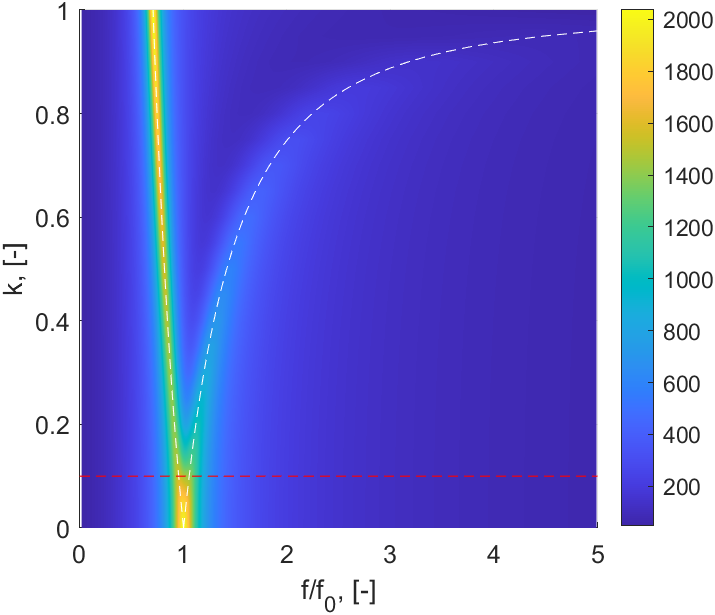}
\caption{Equivalent circuit simulated impedance spectra for coupled resonant inductor circuits showing a) the dual resonant peaks of the frequency spectra and b) coupling coefficient verses frequency dispersion curves, representing the resonant vibrational modes of the system.}
\label{fig:2C_dispersion}
\end{figure}

\subsubsection{Identical Coils}
In the high-frequency case where circuit components of coils 1 and 2 are identical, $R_1 = R_2 = R$, $L_1 = L_2 = L$, and $C_1 = C_2 = C$, such that $M=kL$, and the resonant frequencies $\omega_1 = \omega_2 = \omega_0$, the resonant frequencies and coupling coefficient respectively can be simplified to \cite{arroyo2023displacement},
\begin{align}
\omega_{\pm}' &\approx \sqrt{ \frac{1}{LC\left[1 \pm k \right]} },\label{eqn:f0m1_2}\\
k &\approx \frac{\omega_{-}'{}^2 - \omega_{+}'{}^2}{\omega_{-}'{}^2 + \omega_{+}'{} ^2}.\label{eqn:k2}
\end{align}
Figure~\ref{fig:2C_dispersion}.a shows the frequency spectra of strongly-coupled identical coils and shows how the frequency splitting effect occurs as a function of $k$. The specific impedance equation for a two-coil system cna be found in \cite{arroyo2023displacement}.  Equation~\ref{eqn:f0m1_2} can be used to predict the dispersion of the two resonant frequencies, shown as white dotted lines in Figure~\ref{fig:2C_dispersion}.b, as a function of the coupling coefficient. The system can be thought of as exhibiting independent vibrational modes. Figure~\ref{fig:2C_dispersion}.b also shows a red dotted line representing the dispersion separation threshold; the coupling coefficient of this system above which independent resonant peaks can be resolved.  This threshold is dependent on the Q-factor of the systems and as such lower resistance systems exhibit a sharper, more easily resolvable dispersion at lower coupling coefficients.  There is also a practical upper threshold to the coupling coefficient of a realistic inductively coupled system which is dependent on the geometry of the system and the permeability of the cores used within the coils.

Equation~\ref{eqn:k2} matches the generalised Cohn-Matthaei formulae for coupled resonators derived by Tyurnev 2008 \cite{Tyurnev2008}, and means that the coupling coefficient can be calculated by measurement of the resonant frequencies.  Comparable formula are regularly applied in the modelling of microwave bandpass networks and meta-material design, but the features and properties of coupled resonators are yet to be explored in sensing applications.


\begin{figure}[!h]
\centering
\includegraphics[width=4.0in]{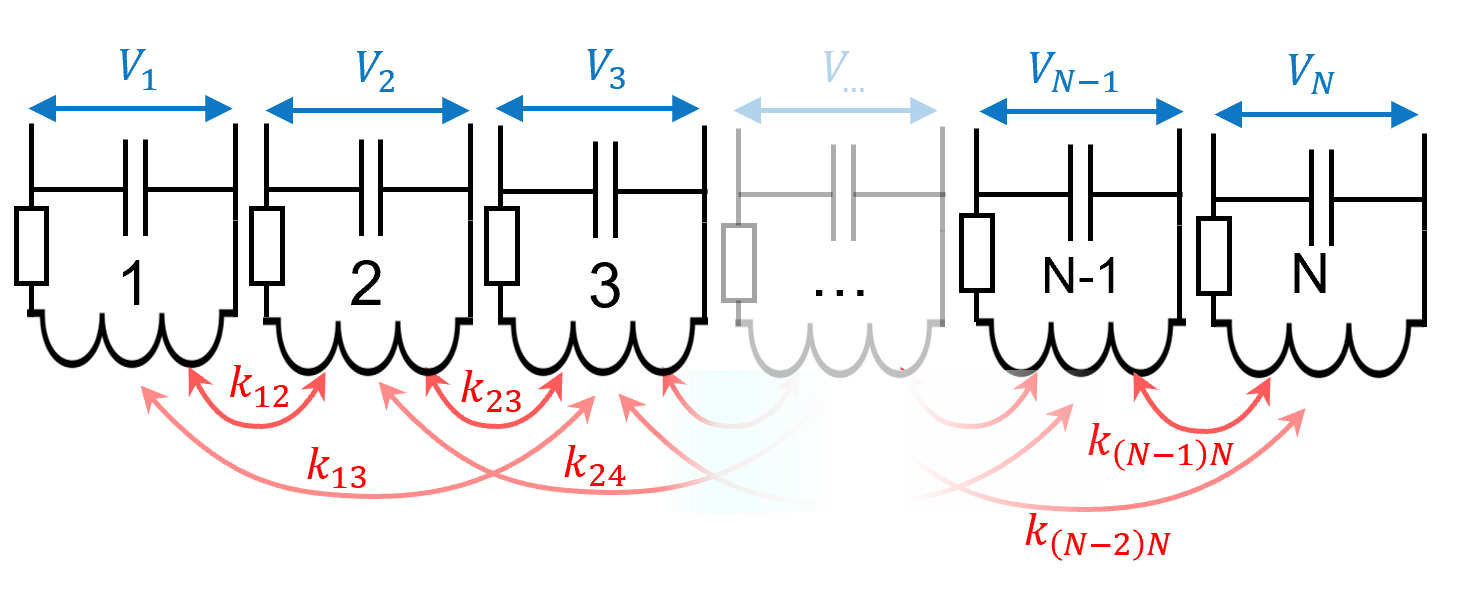}
\caption{Equivalent circuit diagram of an N-dimensional array}
\label{fig:NDcircmodel}
\end{figure}

\subsection{Multi-coil Circuit Model}\label{sec:General Circuit}
Figure~\ref{fig:NDcircmodel} shows an example circuit diagram for a N-dimensional (ND) linear array of resonant coils.  To consider a mutli-coil system (i.e. when the number of coils N > 2) we will evaluate a 1D linear array of three coils before extending to ND. For the 3-coil array, Kirchhoff's laws provide the equations \cite{Zhang2022ANALYSISOA}, 
\begin{align}
V_1 &= I_1 [R_1 + i\omega M_{11}] + i\omega [M_{21}I_{2} + M_{31}I_3], \label{1.1} \\
0 &= I_2 Z_2 +  i\omega[M_{12}I_1 + M_{32}I_3] \label{1.2}, \\
0 &= I_3 Z_3 + i\omega[M_{13}I_1 + M_{23}I_2] \label{1.3},
\end{align}
which can be written in matrix form,
\begin{gather}
\textbf{V} = \mathbb{Z} \textbf{I} \xrightarrow{}
\begin{vmatrix} 
V_{1} \\
0 \\
0
\end{vmatrix} = 
\begin{vmatrix} 
Z_{1} & i\omega M_{12} & i\omega M_{13}\\
i\omega M_{21} & Z_{2} & i\omega M_{23} \\
i\omega M_{31} & i\omega M_{32} & Z_{3} 
\end{vmatrix}
\begin{vmatrix} 
I_{1} \\
I_{2} \\
I_{3}
\end{vmatrix}, \label{eqn:VZI_mat} 
\end{gather} 
where the impedance of the passive circuits ($m>1$) is defined as,
\begin{align}\label{1.4}
\begin{split}
Z_m = R_m + \frac{1}{i\omega C_m} + i\omega M_{mm},
\end{split}
\end{align}
and the mutual inductance between coils is, 
\begin{gather}
\mathbb{M} = 
\begin{vmatrix} \label{eqn:M_matrix3}
M_{11} & M_{12} & M_{13}\\
M_{21} & M_{22} & M_{23} \\
M_{31} & M_{32} & M_{33} 
\end{vmatrix},
\end{gather} 
where $M_{nm} = k_{nm}\sqrt{L_n L_m}$ is the mutual inductance, between coils $n$ and $m$, with the coupling coefficient  $ k_{n,m} = \{ k_{n,m} \in \mathbb{R}, 0 < k < 1 \}$, where $n$ is the index of the excitation circuit, and $m$ is the index of the passive circuit. 
We can therefore define a coupling coefficient matrix,
\begin{gather}
\mathbb{K} = 
\begin{vmatrix}
k_{11} & k_{12} & k_{13}\\
k_{21} & k_{22} & k_{23} \\
k_{31} & k_{32} & k_{33} 
\end{vmatrix}
=
\begin{vmatrix}
1 & k_{12} & k_{13}\\
k_{21} & 1 & k_{23} \\
k_{31} & k_{32} & 1
\end{vmatrix},\label{eqn:k_matrix3}
\end{gather} 
where $k_{nm} = k_{mn}$. When $n=m$, $k_{nn}=1$ and therefore from Equation~\ref{eqn:M_matrix3} $M_{nn}=L_n$. 

Equations~\ref{1.2} and \ref{1.3} can be redefined into a general expression for the current within each circuit,
\begin{align}
\begin{split}
I_m &= \frac{i\omega\left[ \sum_{n \neq m}^{N} M_{nm}I_n \right]}{Z_m}, \label{eqn:I_m}
\end{split}
\end{align}
where $M_{1m}$ and $M_{2m}$ are the nearest and second nearest neighbour mutual inductances experienced by the coil $m$. The voltage experienced by coil $n$ can therefore be defined as,

\begin{align}
\begin{split}
V_n &= I_n \left[R_n + i\omega M_{nn}\right] +  i\omega\left[ \sum_{m \neq n}^{M} M_{mn}I_m \right]. \label{eqn:Vn}
\end{split}
\end{align}

These expressions will remain valid for a coil in a 1D chain for all N and M, representing the number of coils in the chain. 

\subsection{Resonant Modes in ND Arrays}
To formulate an eigenvalue matrix for the coupled inductor system, we must recognise that the system is being measured as a parallel $LC$ resonant circuit.  As such, we must consider the measured admittance, $\mathbb{Y}_{0} = \mathbb{Z}_0^{-1}$, of the coupled system,  
\begin{gather}
\mathbb{Y}_0 \textbf{V} = \textbf{I}.\label{eqn:Y0_V}
\end{gather}
We can define $\mathbb{Y}_{0}$ via the reciprocal rule as, 
\begin{align}
\mathbb{Y}_{0} = \frac{1}{\mathbb{Z}_0} &= \frac{1}{\mathbb{Z}_L} + \frac{1}{\mathbb{Z}_c} 
,\\
&= \mathbb{Z}_L^{-1} + {i\omega \mathbb{C}} 
,\label{eqn:Y0}
\end{align}
where $\mathbb{C}$ is the $N\times N$ parallel capacitance matrix, 
\begin{gather}
\mathbb{C} =
\begin{vmatrix} 
C_1 & 0 & \cdots & 0\\
0 & C_2 & \cdots & 0 \\
\vdots & \vdots & \ddots & \vdots \\
0 & 0 & \cdots & C_N
\end{vmatrix}, \label{eqn:Cmat}
\end{gather}
with $C_n$ representing the parallel capacitances in inductor circuit, $n$. 
$\mathbb{Z}_L$ is the $N\times N$ coupled inductive impedance matrix, which can be approximated in the high-frequency case ($R\ll \omega L$) to,
\begin{gather}
\mathbb{Z}_L \approx i\omega \mathbb{M} = i\omega
\begin{vmatrix} 
M_{11} & M_{12} & \cdots & M_{1N}\\
M_{21} & M_{22} & \cdots & M_{2N} \\
\vdots & \vdots & \ddots & \vdots \\
M_{N1} &  M_{N2} & \cdots & M_{NN} 
\end{vmatrix},
\end{gather}
where $M_{nm} = k_{nm}\sqrt{L_n L_m}$, and where $k$ and $L$ have their usual meaning as the coupling coefficient and inductance respectively.  
We can write Equation~\ref{eqn:Y0} as,
\begin{align}
\mathbb{Y}_{0} &= \frac{1}{i\omega \mathbb{M}} + {i\omega \mathbb{C}}, \\
 &= \frac{1}{i\omega}\left( \mathbb{M}^{-1} - \omega^2 \mathbb{C} \right).
\label{eqn:Apdx_Y0_2}
\end{align}
Inputting Equation~\ref{eqn:Apdx_Y0_2} into Equation~\ref{eqn:Y0_V} and rearranging gives,
\begin{align}
\left[ \mathbb{C}\left( \mathbb{\Omega} - \omega^2 \mathbb{1} \right) \right]\textbf{V}= {i\omega }\textbf{I}, \label{eqn:Apdx_Y0_3}
\end{align}
where $\mathbb{\Omega}$ is the characteristic eigen-matrix with the form,
\begin{equation}
\mathbb{\Omega} = \left[\mathbf{C}\mathbb{M}\mathbf{C}^T\right]^{-1} = \Omega\mathbb{K}^{-1}\Omega^T,
\end{equation}
$\mathbf{C}$ and $\Omega$ are $N\times N$ diagonal matrices with elements $\sqrt{C_n}$ and $\omega_n$ respectively, and
\begin{gather}
\Omega = 
\begin{vmatrix} 
\omega_1 & 0 & \cdots & 0\\
0 & \omega_2 & \cdots & 0\\
\vdots & \vdots & \ddots & \vdots \\
0 & 0 & \cdots & \omega_N
\end{vmatrix}
,\label{eqn:LL}
\end{gather}
 $\omega_n = \sqrt{L_nC_n}^{-1}$ is the natural resonant frequency of an uncoupled coil $n$. The inverse coupling coefficient matrix $\mathbb{K}^{-1}$ can be calculated using the standard inverse matrix identity ($\mathbb{K}^{-1} = \nicefrac{Adj.\mathbb{K}}{|\mathbb{K}|}$).

Given that the impedance $Z_0$ is a maximum (infinite in the idealised cases) at resonance, we can conclude that the admittance $Y_0$ will be a minimum at resonance (zero in idealised case), such that the current $\mathbf{I}$ becomes zero at resonance. Equation~\ref{eqn:Apdx_Y0_3} can therefore be rearranged into the form,
\begin{align}
\left( \mathbb{\Omega} - \omega_i^2 \mathbb{1} \right) \vec{v}_i = 0, \label{eqn:Apdx_Y0_4}
\end{align}
which takes the form of a traditional eigen-equation, where $\vec{v}_i$ is the $i^{th}$ voltage eigenvector, and $\omega_i^2$ is the eigenvalue of the characteristic matrix, $\mathbb{\Omega}$, and satisfies the expression,
\begin{align}
\mathbb{\Omega}\vec{v}_i = \omega_i^2 \vec{v}_i. \label{eqn:Apdx_eigEq_1}
\end{align}
We can therefore calculate the eigenvalues and vectors of the coupled system by solving,
\begin{align}
    \det{\left( \mathbb{\Omega} - \omega_i^2 \mathbb{1} \right)} &= 0. \label{eqn:Apdx_eigEq_2}
\end{align}
With all electronic components in the system known, and if the coupling coefficients, $k_{nm}$, between array elements can be estimated, we can predict the eigenvalues (resonant frequencies) of any ND-coupled system.  From this matrix analysis, it is clear that an overcoupled system will exhibit as many resonant modes as there are array elements.

\begin{figure}[!t]
\centering
\includegraphics[width=6.4in]{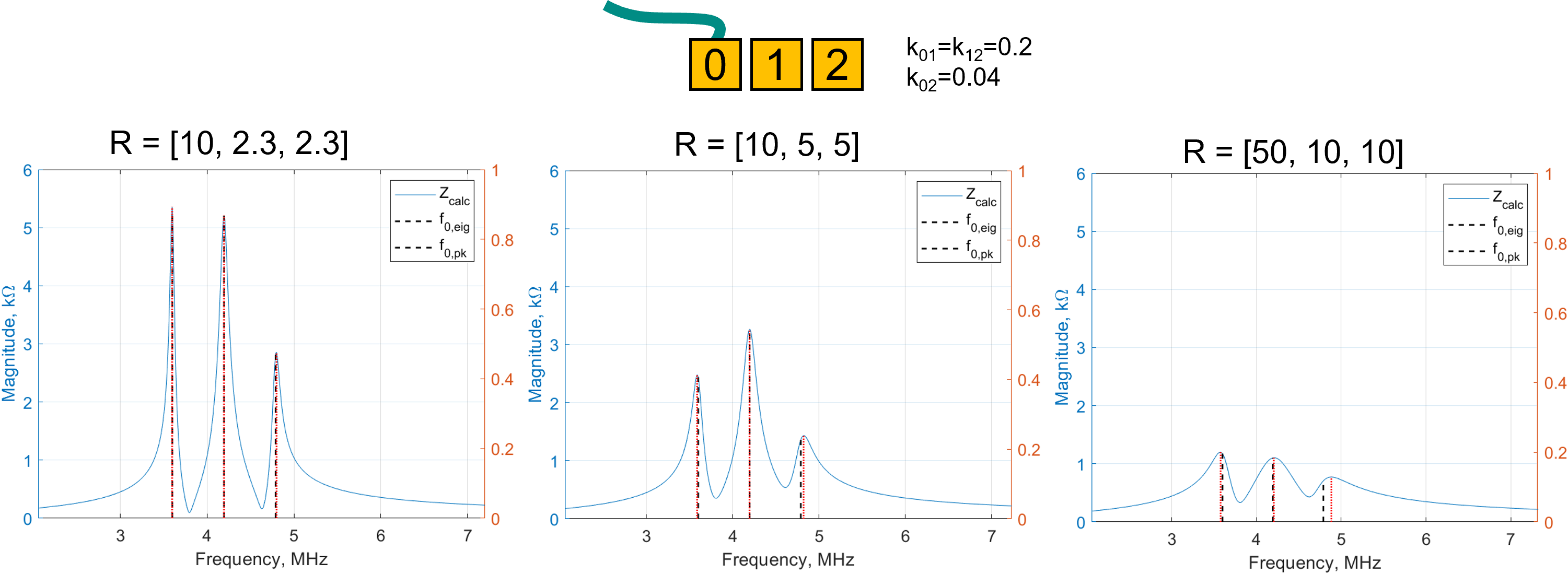}
\caption{Circuit model predicted impedance magnitude profiles of a 3-coil linear array systems of identical coil inductance ($10~\mu H$) and capacitance ($150~pF$) with varying series resistance, $R$, showing the resonant damping effect and disparity between the eigen-matrix calculated resonant frequencies (black dotted lines) and the circuit predicted peak frequencies (red solid lines).}
\label{fig:exampleZf_modes}
\end{figure}

Figure~\ref{fig:exampleZf_modes} shows the the equivalent circuit theory impedance magnitude spectra calculated using equations~\ref{eqn:I_m}-\ref{eqn:Vn} for a linear 3-coil array with increasing resistance, demonstrating the difference between the eigenmode predicted resonant frequencies and the actual peak impedance frequencies as the high Q-factor assumption breaks down.

\section{Materials \& Methods}
The significance of the above impedance formulation and eigenvalue prediction is evaluated in the following section for some example array configurations; qualitatively comparing circuit model predicted impedance profiles with 2D finite element model (2DFEM) simulated results and experimental measurements.

\subsection{2D Finite Element Simulation}\label{FEM}
Two-dimensional (2D) Finite element models (FEM) were created in COMSOL 6.1 using the AC/DC module. Models were developed using Magnetic Fields and Electrical Circuit studies to simulate the equivalent circuit of resonating coils and were evaluated in the frequency domain. The models simulated infinitely long (extending into the page) coil arrays of height $10~mm$, core radius $2~mm$, with homogenised multi-turn coil windings of thickness $0.6~mm$.  The core material was ferrite, with a relative permeability $\mu_r=25$.  Each coil was capacitively loaded with a capacitor of $50~pF$ and a series resistance of $200~\Omega$.  Due to the 2D nature of the simulation, the inductances of the coil array elements were unrealistically high, but the prediction of the relative resonances in the arrays remains valid.

From these simulations, the electrical impedance spectra, $Z(f)$, the array element voltages, $V_n$, and the spatial distribution of the magnetic flux density, $B(f,x,y)$, can be evaluated as a function of the excitation frequency. 

\subsection{Experimental Validation}\label{Exp}
Quantitative experimental validation is performed by evaluating the impedance spectra of a simple linear array of $2~mm$ outer radius inductors, of inductance $10~\mu H$ and circuit capacitance of $1.7~nF$, giving each element a natural resonant frequency of $1.2~\mbox{MHz}$. Each coil has a ferrite core (Grade 61, FairRite) of diameter $1~mm$, and height $10~mm$.  The impedance spectra of the linear array was measured using an Impedance Analyser (Trewmac TE3001) between $0.5-1.5~\mbox{MHz}$ across each of the unique coil array elements. The voltages across each array element were measured at each resonant frequency using a Picoscope 4824A USB oscilloscope, while a SonEMAT Howland Current Source was used to convert a voltage input waveform from a RS Pro RSDG 830 Arbitrary Function Generator into a controlled alternating current supplied to the central array element.

\section{Results \& Discussion}
A number of specific array configurations are considered in this study, including identical resonator arrays in linear and close-packed arrangements, as well as disparate resonant arrays where each element exhibits a unique resonant frequency achieved via capacitive tuning. 
In each case, circuit models, 2D FE simulated and experimental results are compared and contrasted with specific attention to the qualitative phenomenon and features of the array's impedance and magnetic flux density.  
While the theoretical models presented in section~\ref{sec:theory} can be applied to any physical configuration of coils, even those with disparate circuit parameters and resonances, in this paper we focus on arrays of nominally identical coplanar inductors as an example validation case, i.e. with identical coil properties and capacitive loads such that each coil element before installation in the array has identical resonant frequencies $f_n = f_m = f_0$.

\subsection{Three-Coil Arrays}
\begin{figure}[!t]
\centering
\includegraphics[width=0.6\textwidth]{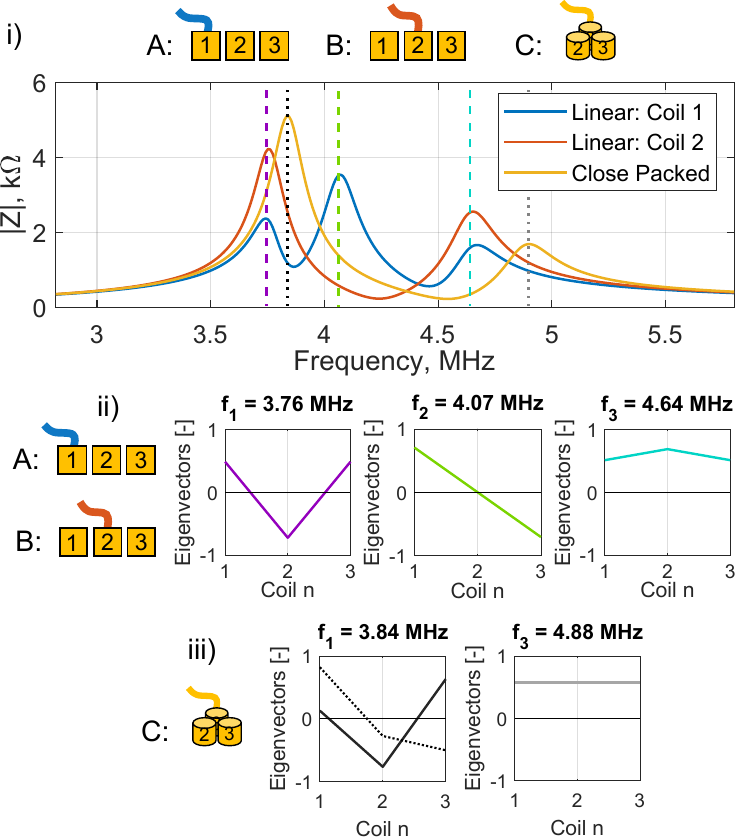}\\
\caption{Predicted impedance spectra for three 3-coil over-coupled arrays configurations of identical coil inductance ($10~\mu H$), capacitance ($150~pF$), series resistance ($10~\Omega$) and coupling coefficient ($k = 0.14$) showing i) impedance magnitude spectra $|Z(f)|$ for three coil measurement configurations; Linear (A and B), and iii) Close-packed (C) arrays. Plots in blue and red represent measurements of a linear array from the end and centre coils respectively. The yellow curve shows the measurement for a close-packed coil configuration (C). Dotted lines correspond to the theoretically predicted resonant eigen-frequencies with colours corresponding to the predicted eigen-modes shown in the ii) and iii) for linear and close-packed arrays respectively.}
\label{fig:exampleZf}
\end{figure}

Examples of predicted impedance profiles, $Z_0(f)$, generated using the general circuit model in Section~\ref{sec:General Circuit} are shown in Figure~\ref{fig:exampleZf}. These are compared to the predicted eigen-frequencies, $f_n$ (dashed vertical lines), from equation~\ref{eqn:Apdx_eigEq_2} for three different 3-coil configurations; a linear array measured at the end and central coil respectively and a close-packed array. 
Each coil is an identical resonator with inductance $L=10~\mu H$, capacitance $C=150~pF$, and resistance $R=10~\Omega$.

Figure~\ref{fig:exampleZf} also shows the eigenvectors (representing the array excitation modes) found from Equation~\ref{eqn:Apdx_eigEq_1}, providing information on which elements are excited, and to what extent, at each resonant frequency, as well as which elements are unexcited (i.e. array nodes). 

The simulated results show that although the eigen-matrix predicts identical resonant frequencies and modes regardless of which element is measured, only the impedance profile of measurements made on the end coil exhibit three distinct resonant peaks.  
This is due to the measurement asymmetry enabling three unique modes, while the symmetry of the linear array measured at the central coil leads to only 2 unique resonant peaks. 
This is shown by the second mode ($f_2$) where the central coil has an eigenvalue of 0.

The mode-shapes shown in Figure~\ref{fig:exampleZf}.ii and iii represent the excitation states of each element at each resonant frequency.  Both linear arrays (Figure~\ref{fig:exampleZf}.a \& b) exhibit the same eigen-modes and resonant frequencies, however when measuring from the central coil, only two resonant peaks are observed with the 2nd resonant frequency corresponding to a mode where the central coil exhibits a node state in the eigen-modes.  In comparison, Figure~\ref{fig:exampleZf}.iii shows that a close-packed 3-coil array experiences different resonant frequencies to the linear array configuration, and there exists two simultaneous modes at the same resonant frequency. This is believed to correspond to the equal chance of coil 1 being in phase with either coil 2 or 3.

This eigen-mode analysis has the potential to be extremely powerful for exciting arrays in specific spatial coil configurations by selecting appropriate frequencies. 
For example, in a linear array, the lowest resonant frequency corresponds to an alternating eigenmode between coil elements corresponding to a spatially alternating magnetic field direction. Conversely, exciting the array at the highest frequency will excite a mode whereby all elements are in phase such that the array behaves like a single larger coil.

\subsubsection{2D FEM Analysis}
\begin{figure}[!t]
\centering
\includegraphics[width=0.9\textwidth]{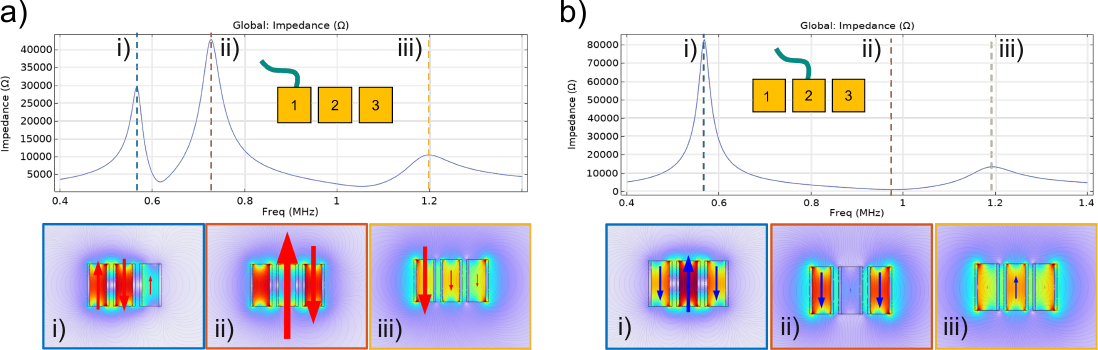}
\caption{2D FEM impedance spectra for two linear 3-coil over-coupled arrays configurations of identical resonance showing impedance magnitude $|Z|$ spectra and the spatial distribution of the magnetic flux density at each resonant frequency (i-iii). Showing a) linear array measured from active end coil, and b) measured from an active central coil.}
\label{fig:3coilFE_Zf}
\end{figure}

Figure~\ref{fig:3coilFE_Zf} shows the impedance spectra predicted using a 2D FEM of a linear 3-coil array as measured from an active (a) end and (b) central coil.  The results show the same resonant splitting phenomena predicted by the circuit model (see Figure~\ref{fig:exampleZf}), with 3 unique modes predicted for the end-excited coil, and only 2 unique resonant peaks for the centrally-excited coil.  The FEM predicts the same patterns of each resonant peak as the circuit model, with key features being; 1) the first and second peaks occurring in close proximity in the frequency domain, 2) the second peak having the highest amplitude, and 3) the third resonant peak having the lowest amplitude resonance mode. The centrally-excited array also exhibits the same resonant spectra as predicted by the circuit theory, with the lowest frequency having the greatest prominence.

The spatial distributions of the magnetic flux density on the bottom tiles in Figure~\ref{fig:3coilFE_Zf} incorporate arrows that show the magnitude and direction of the magnetic flux density in the middle of each coil element at each resonant frequency of the array. This analysis shows the spatial patterns of the magnetic flux in these co-planar arrays is shared between coil elements in different mode-shapes mimicking the eigenvectors predicted by the circuit model.  

\subsection{Five-Coil Linear Array}
\begin{figure}[!b]
    \centering
            a)\\
            \includegraphics[clip, trim = 0cm 0cm 0cm 0cm, width =0.39\textwidth]{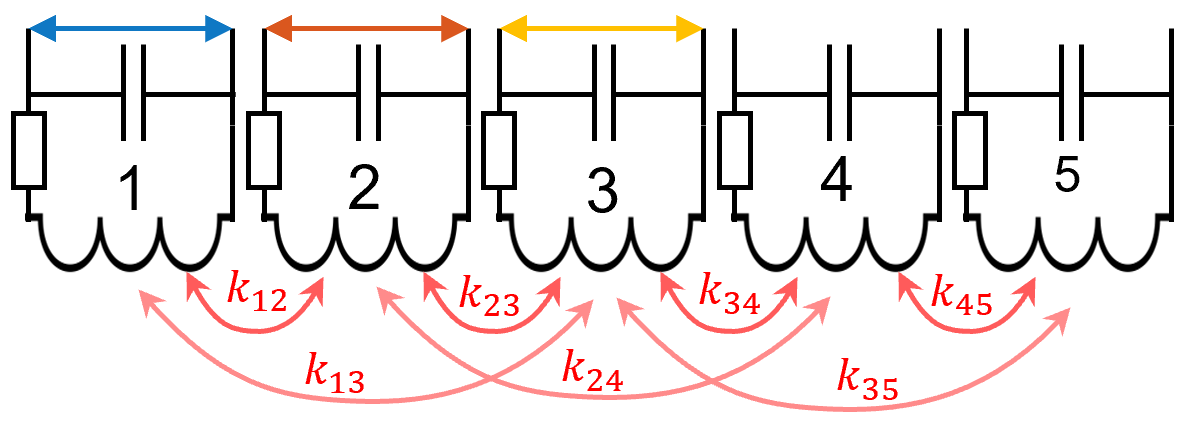}\\ 
            b)\\
            \includegraphics[width =0.49\textwidth]{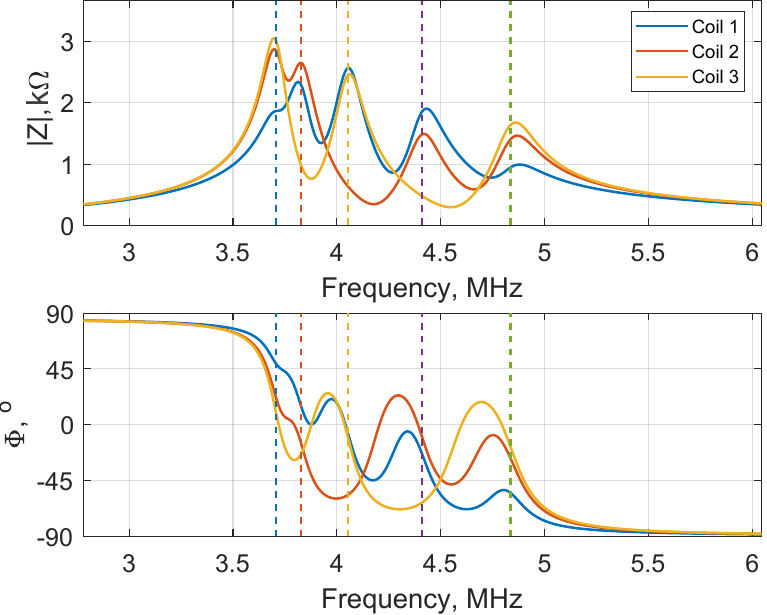} 
            
             c) \\
            \includegraphics[clip, trim = 2cm 2.5cm 2cm 2.5cm, width = 0.9\textwidth]{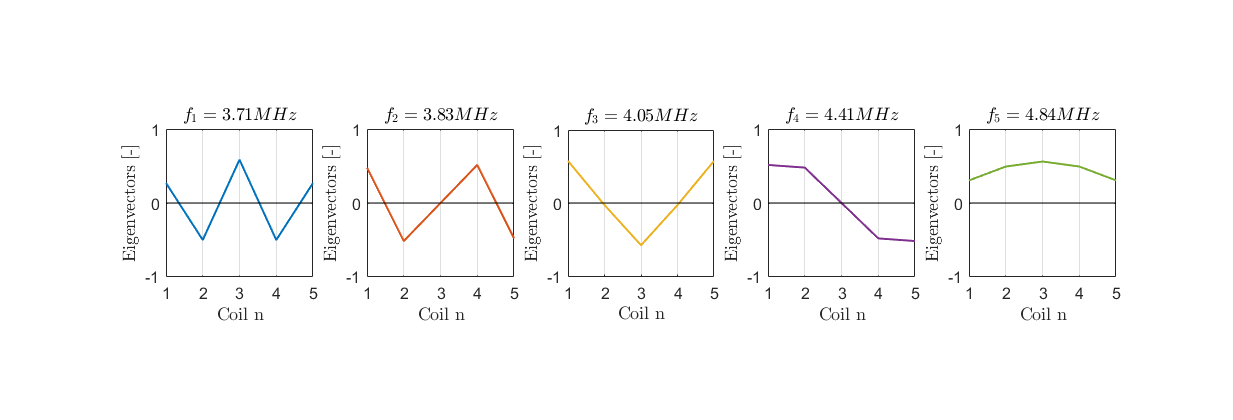}
    \caption{Circuit model prediction - Five identically resonating coils in a linear array, showing; a) linear resonant array configuration (blue = coil 1 active, red = coil 2 active, yello = coil 3 active), b) magnitude $|Z|$ and phase $\Phi$ of impedance spectra as measured across coils 1-3, c) predicted eigenvector modeshapes of the array coil elements at the eigen-matrix predicted resonant frequencies $f_n$.}
    \label{fig:5coil_preds}
\end{figure}

Figure~\ref{fig:5coil_preds} shows how the addition of further array elements leads to additional resonant modes.  The impedance spectra show how the number of peaks changes depending on which coil is used to excite the array, in a similar way to the 3-coil array above.

\begin{figure}[!t]
    \centering
            a) \\
            \includegraphics[width = 0.6\textwidth]{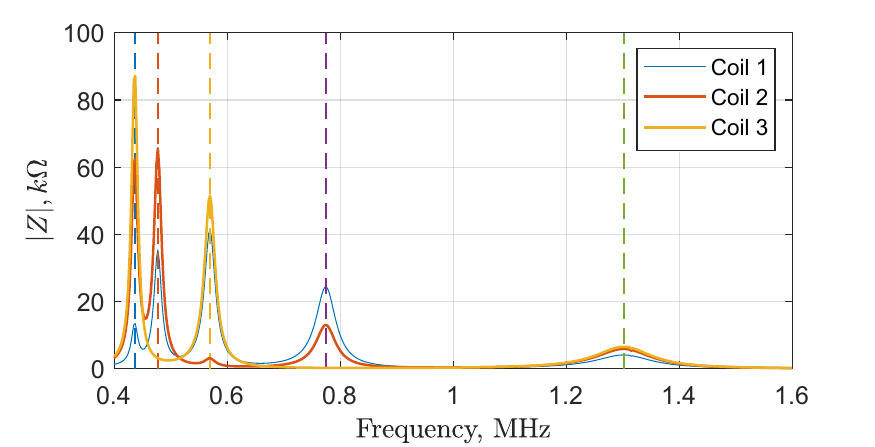}\\
            b) \\
            \includegraphics[clip, trim = 2cm 2.5cm 2cm 2.5cm,width=\textwidth]{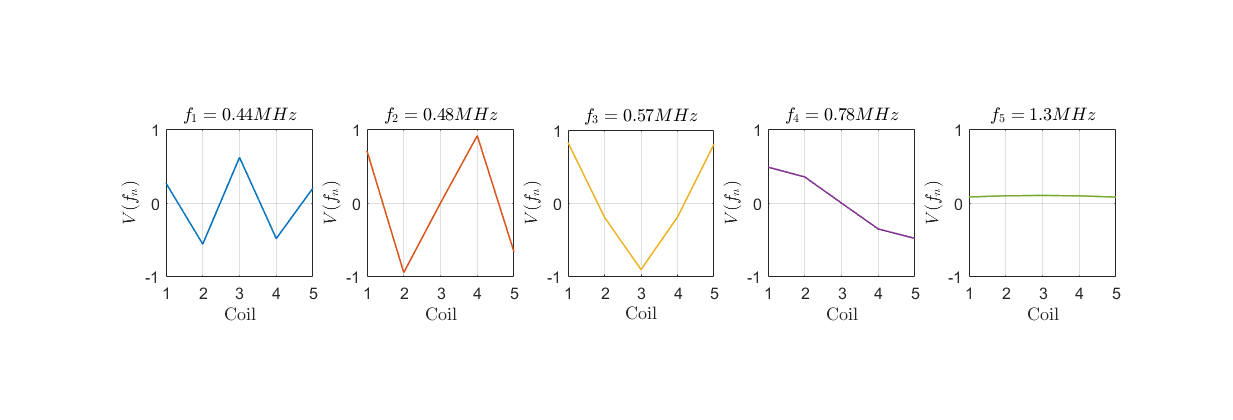}\\
            c) \\
            \includegraphics[width=0.19\textwidth]{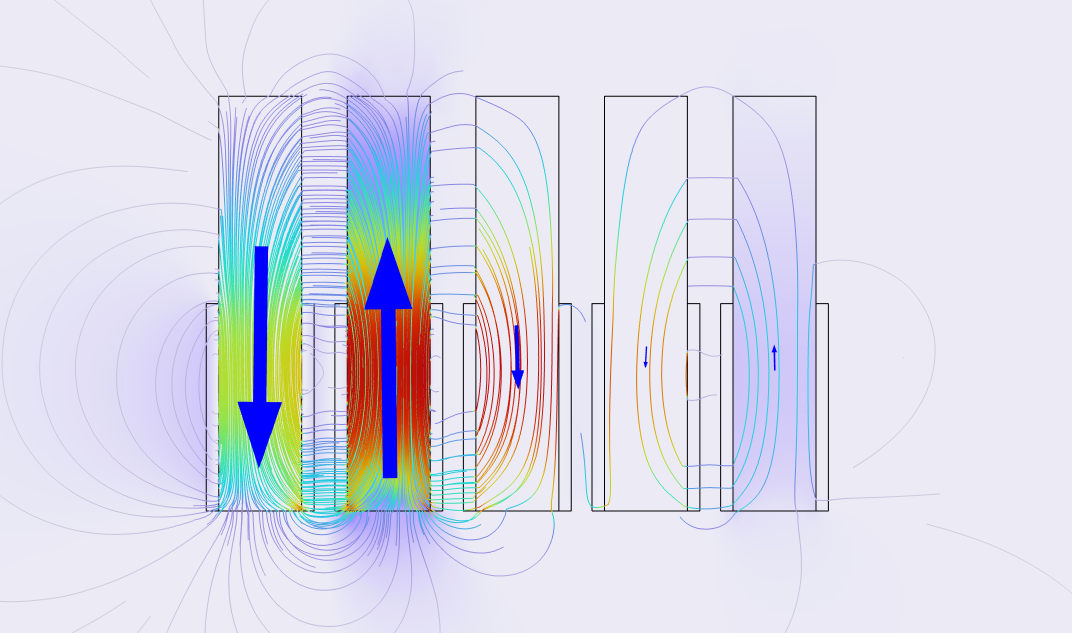}
            \includegraphics[width=0.19\textwidth]{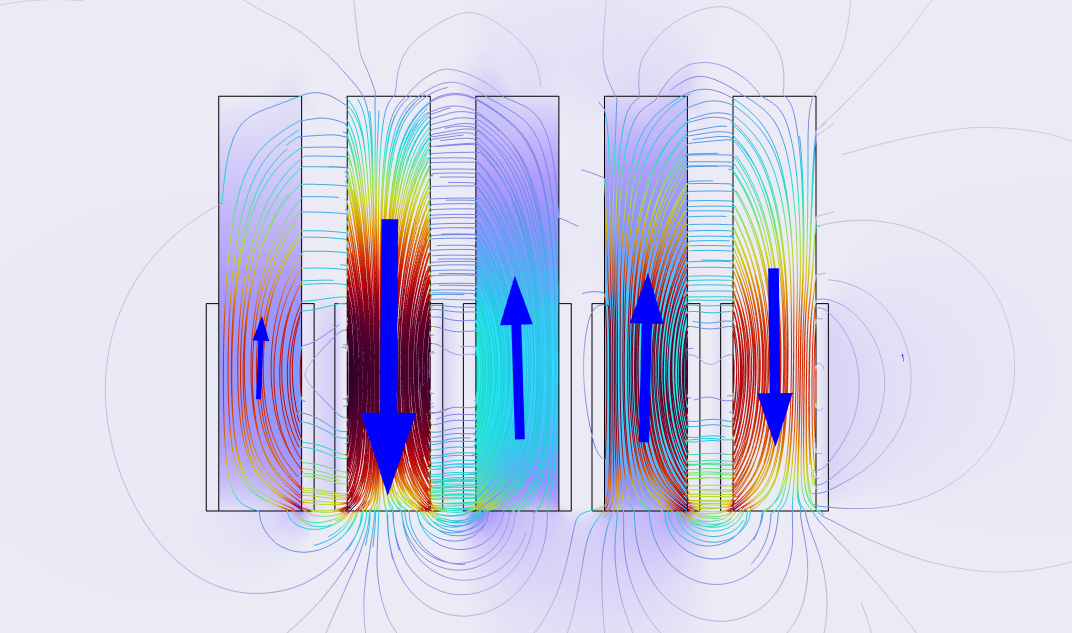}
            \includegraphics[width=0.19\textwidth]{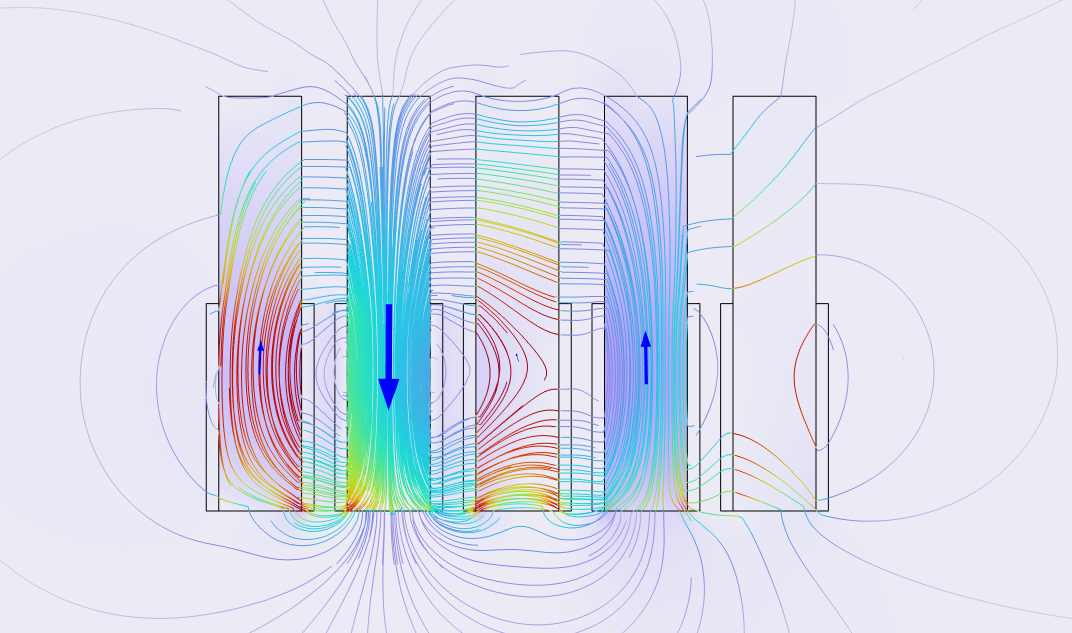}
            \includegraphics[width=0.19\textwidth]{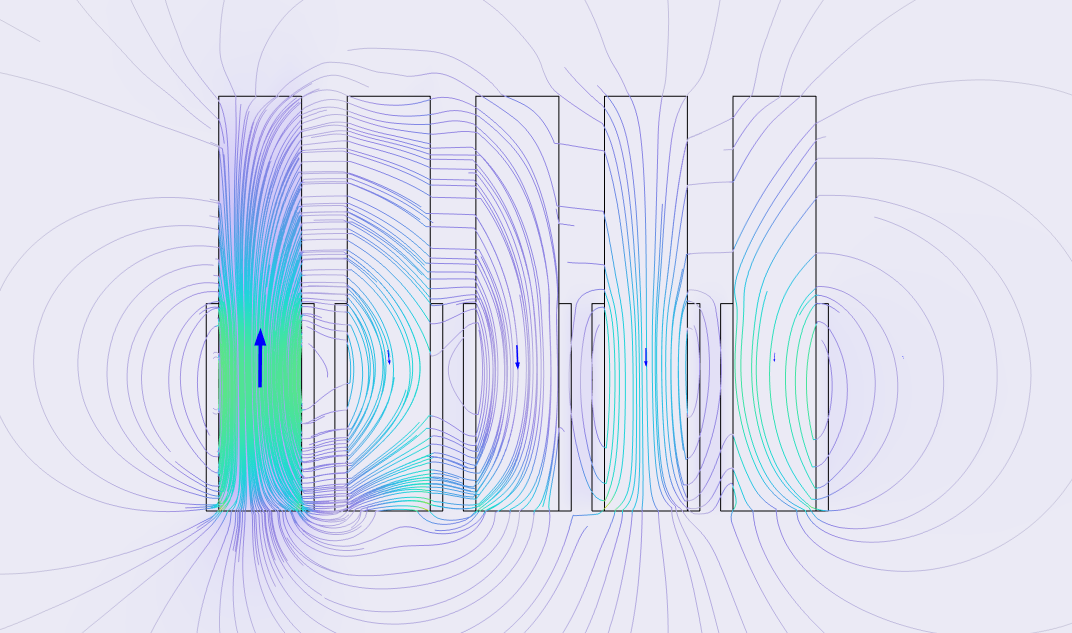}
            \includegraphics[width=0.19\textwidth]{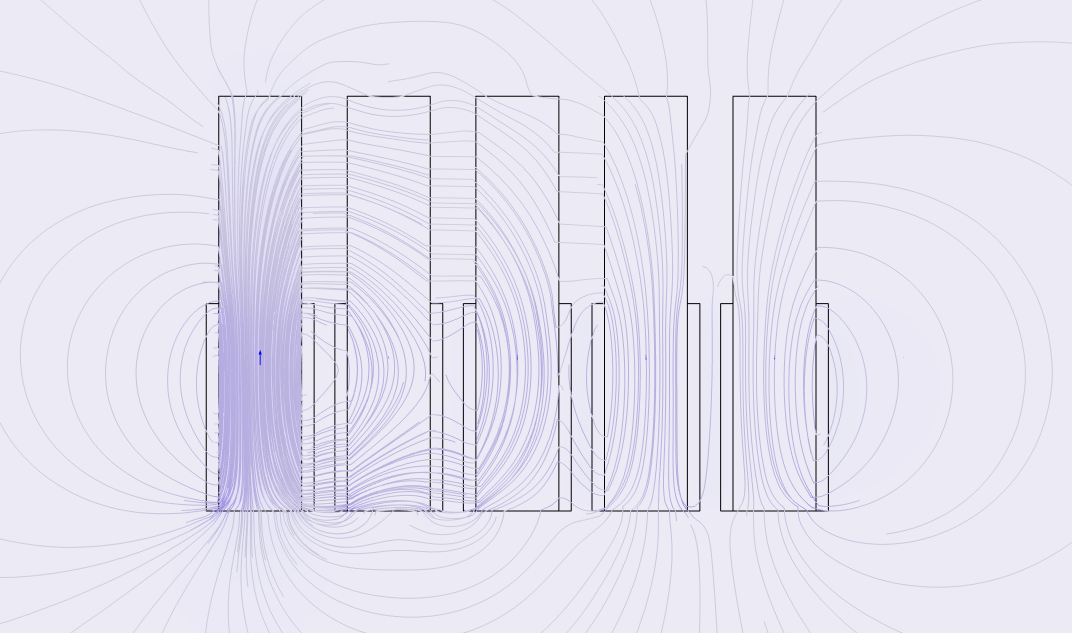}
    \caption{2D FE five-coil linear resonant array spectra - Measured across LC resonator element 1, 2 \& 3 (a), simulated array element voltages (b) and magnetic flux density at the resonant frequencies of the simulated array (c) when excited by coil element 1.}
    \label{fig:5coil_mods}
\end{figure}

Figure~\ref{fig:5coil_mods}.a shows the impedance spectra predicted using a 2D FEM of a linear five-coil array as measured from coils 1 (blue), 2 (red) and 3 (yellow).  
The results show the same resonant splitting phenomena predicted by the circuit model (see Figure~\ref{fig:5coil_preds}), with five unique modes predicted for the end-excited array, and only three unique resonant peaks for the centrally-excited array.  
The FEM predicts the same patterns of each resonant peak as the circuit model, with key features being; 1) the first (i) and second (ii) peaks occurring in close proximity in the frequency domain, 2) the third (iii) peak being the most prominent, and 3) the fifth (v) resonant peak being the least prominent resonance mode. 
The centrally-excited array also exhibits the same resonant spectra as predicted by the circuit theory, with the lowest frequency having the greatest prominence (yellow lines in Figures~\ref{fig:5coil_preds} and \ref{fig:5coil_mods}).  These resonant rules are consistent with the three-coil case.

Figure~\ref{fig:5coil_mods}.b shows the predicted voltage predictions for for each coil element at each resonant frequency. The voltage distributions between coils corresponds directly to the equivalent eigen-mode shapes predicted in Figure~\ref{fig:5coil_mods}.

Figure~\ref{fig:5coil_mods}.c show the magnitude and direction of the magnetic flux density in the middle of each coil element at each resonant frequency of the simulated array. 
This analysis shows the spatial patterns of the magnetic flux in these co-planar arrays as it is shared between coil elements in different mode-shapes mimicking the eigenvectors predicted by the circuit model.   

\subsection{Experimental Validation}
\begin{figure}[t!]
    \centering
            a)\\
            \includegraphics[clip, trim = 1.5cm 0.2cm 2.5cm 0.5cm,width =0.38\textwidth]{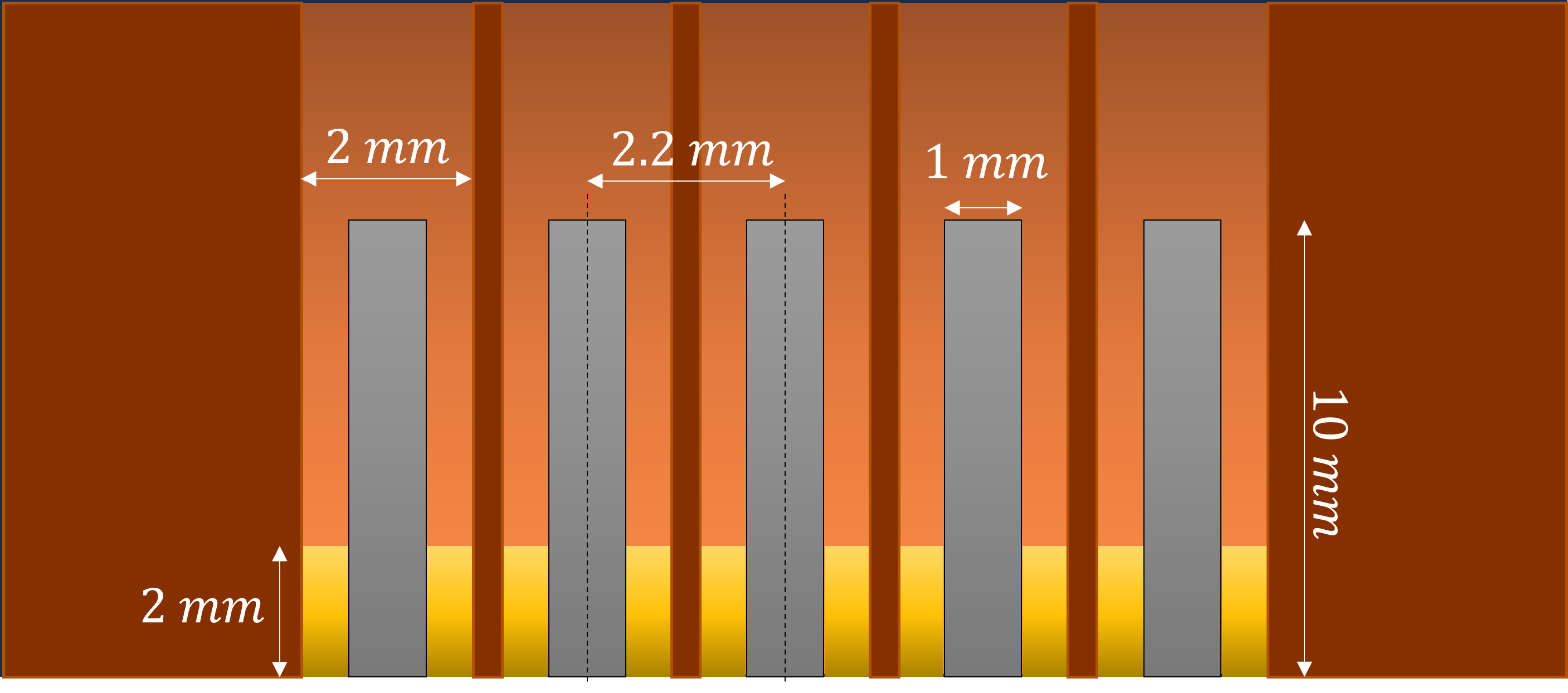}
            \includegraphics[clip, trim = 25cm 40cm 35cm 15cm,width =0.3\textwidth]{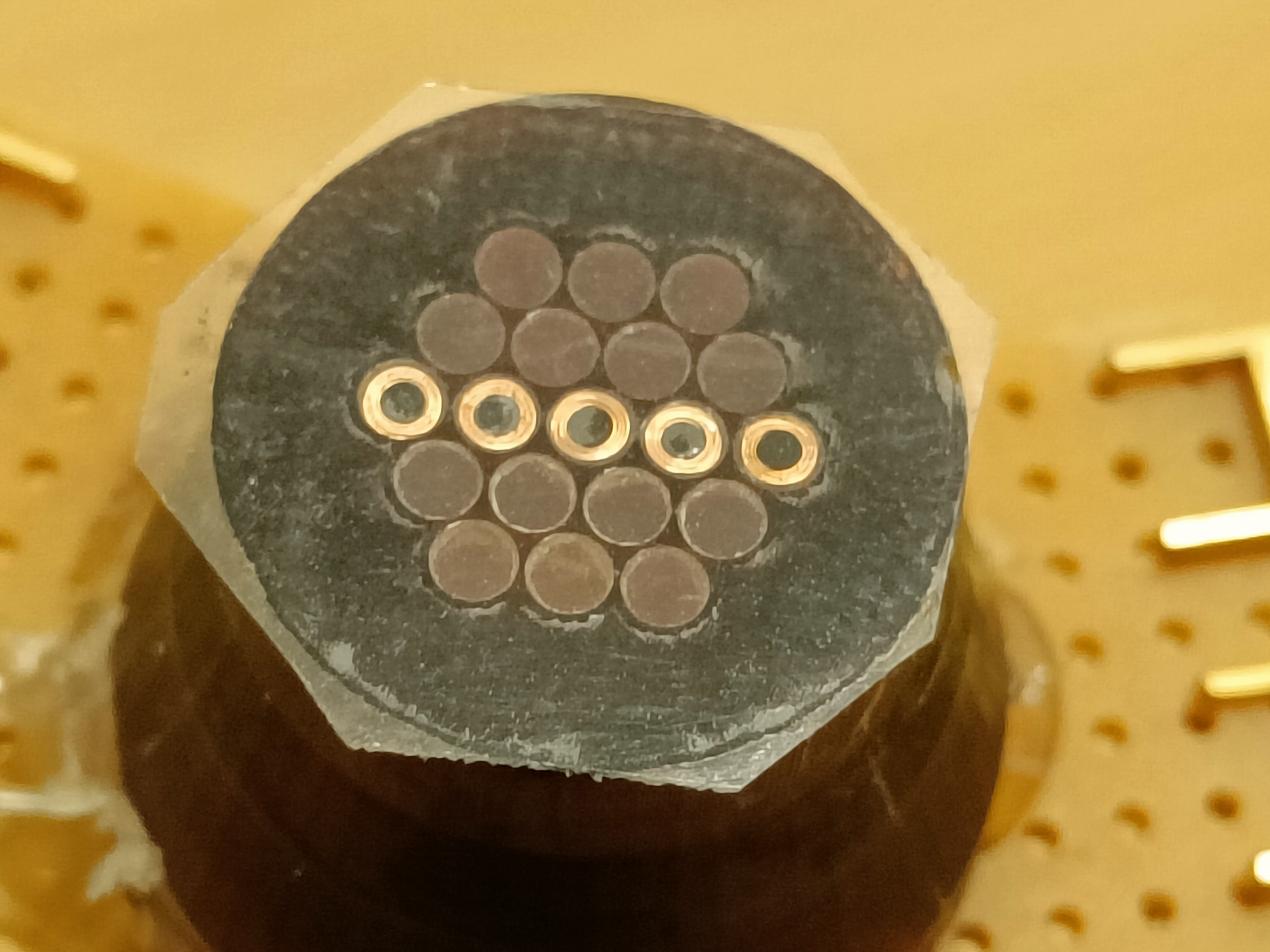}
            \includegraphics[clip, trim = 35cm 40cm 25cm 15cm,width =0.3\textwidth]{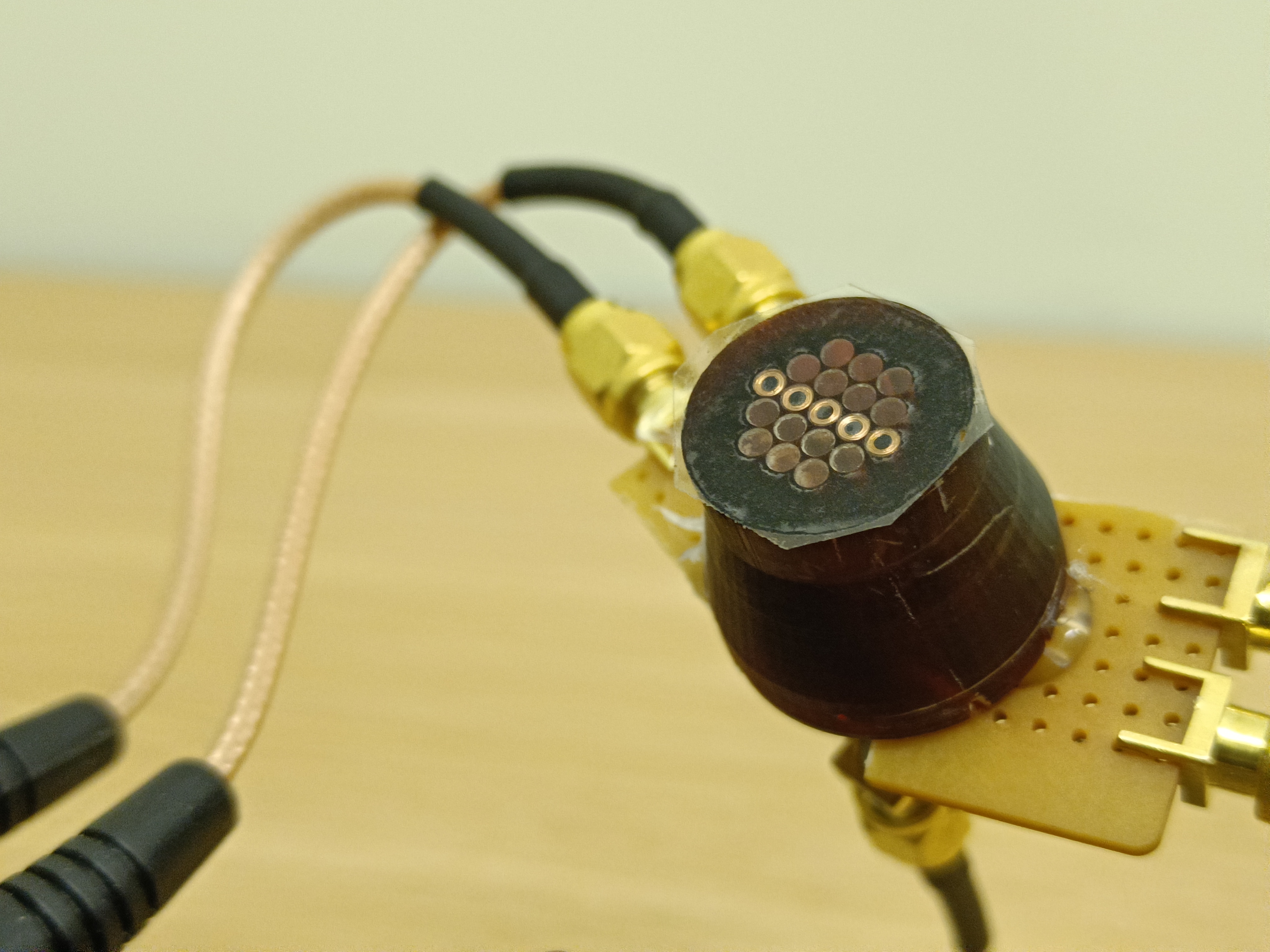}\\
            \smallskip
            b)\\
            i) Experimental \hspace{5cm} ii) Circuit Theory \\
            \includegraphics[clip, trim = 2.0cm 9cm 3cm 9.8cm,width =0.49\textwidth]{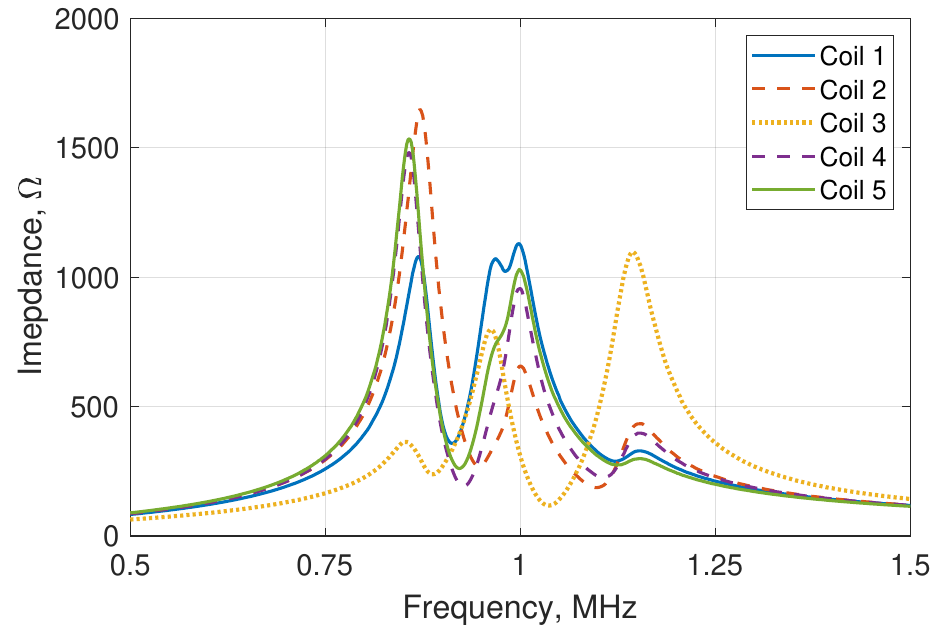}
            \includegraphics[clip, trim = 2.0cm 9cm 3cm 9.8cm,width =0.49\textwidth]{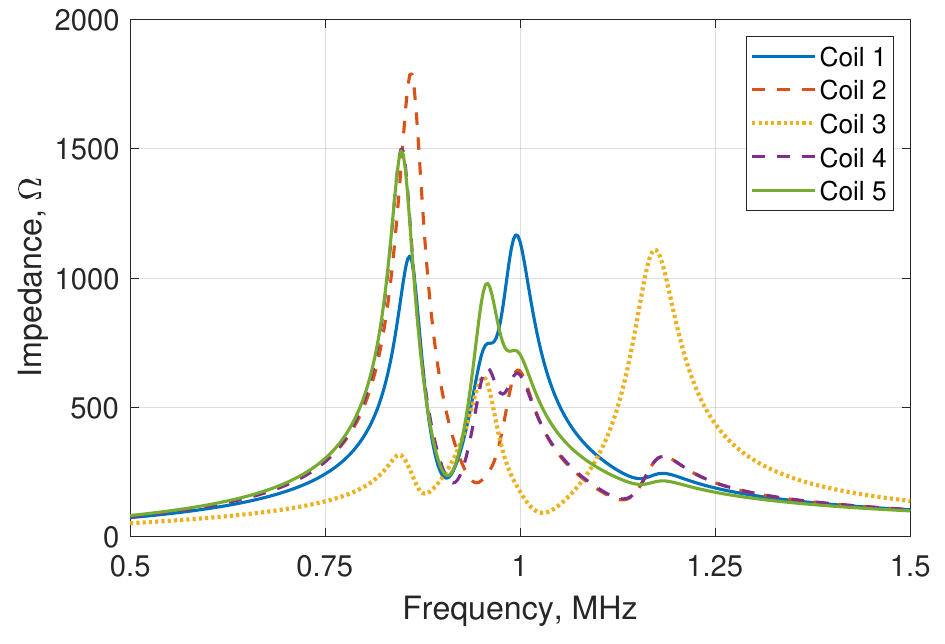}\\
        \smallskip
        c)\\
        \smallskip
        \includegraphics[clip, trim = 0cm 0cm -0.5cm 0cm,width=\textwidth]{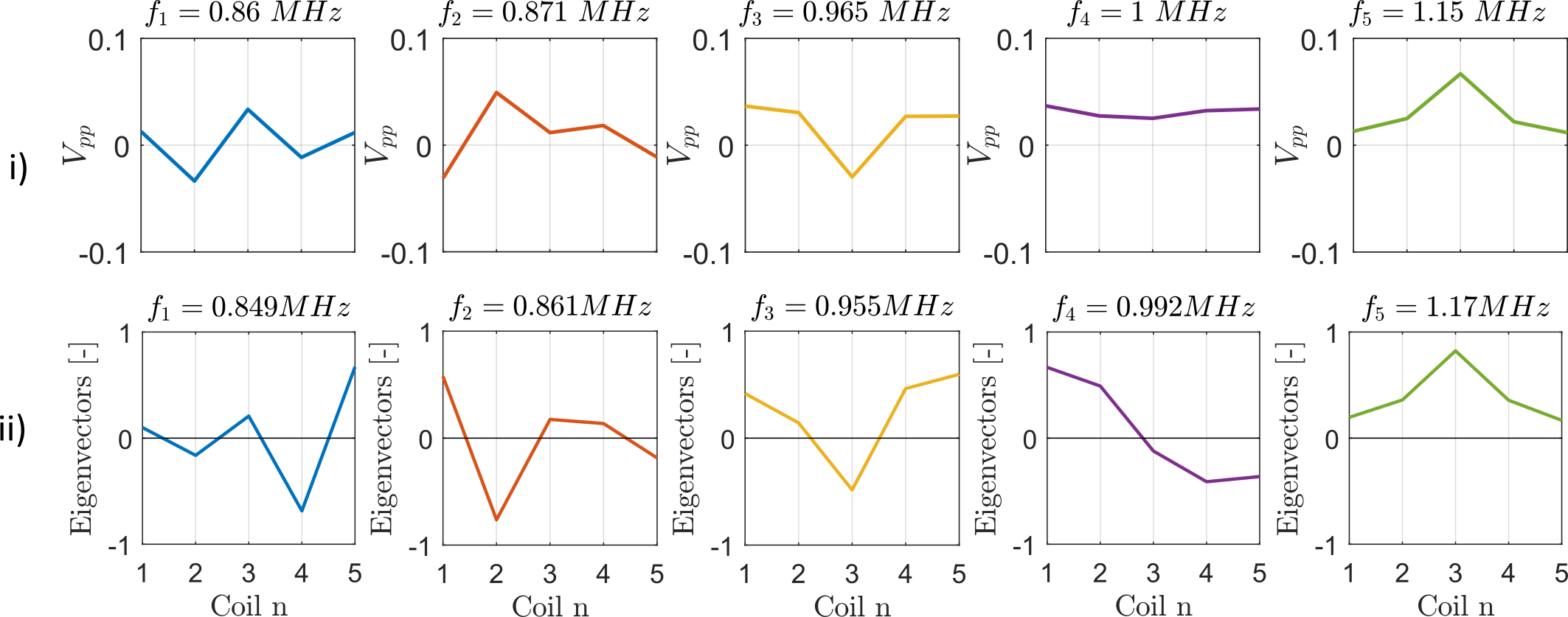}\\
    \caption{Experimental and Circuit Model Comparison for a five-coil array, showing a) the experimental five-coil array configuration and, b) the impedance spectra $|Z(f)|$ showing the experimental (i) and circuit model predicted spectra (ii) when the active coil is changed between each coil element in the configuration. c) Shows i) the experimentally measured voltage outputs of each array element at each resonant frequency (excitation coil independent), compared to the ii) predicted eigenmodes from circuit theory.}
    \label{fig:five-coil-exp}
\end{figure}

In order to experimentally validate the predictions of both the circuit theory and the FEM, an experimental five-coil linear array was manufactured. The array consisted of five identical cylindrical coils 
of free-space inductance $L = 10 \pm 0.1 ~\mu H$ and dimensions shown in Figure~\ref{fig:five-coil-exp}.a (all dimensions are given with a measurement error of $\pm 0.1 mm$).  Each coil was positioned flush with one end of a ferrite (Grade 61, FairRite) rod of diameter $1~mm$ and length $10~mm$, and each was housed in a linear configuration in a SLA 3D-printed plastic array housing, containing five $2~mm$ diameter holes separated by a $200~\mu m$ wall. 

Each array element was connected in parallel to a $1.4 \pm 0.1~nF$ surface mount capacitor, a surface mount SMA connector, and a $0.1 m$ SMA to BNC coaxial cable. The total capacitance and resistance of each array element were measured using a digital multi-meter and displayed in Table~\ref{tab:5coilvals}.


A TREWMAC TE3001 Impedance Analyser was used to measure the resulting impedance spectra of each element within the array and the results are plotted in Figure~\ref{fig:five-coil-exp}.b (left).  
In comparison to the idealised symmetric theoretical impedance spectra shown in Figure~\ref{fig:5coil_preds}.b for coil 1, the experimental results do not show the five distinct resonant peaks predicted by the circuit theory of an identical resonator array. 
They instead show a maximum of four identifiable peaks with the central two peaks being so close in proximity that they are rendered indistinguishable when measured from coil 5 (which should have an identical spectra to coil 1).  
This comparison between coil 1 and 5 spectra demonstrates the asymmetry of the physical array caused by natural variability between the circuit components, and inconsistencies in the positioning of the elements within the array. However, this asymmetry also helps to explain why the spectra are not identical to the idealised theoretical case shown in Figure~\ref{fig:5coil_preds}.b.

The right-hand spectra shown in Figure~\ref{fig:five-coil-exp}.b are the circuit theory predicted impedance spectra for an asymmetric array, specifically produced via manually applying asymmetry to the inductance of each array element with values used shown in Table~\ref{tab:5coilvals}. 

\begin{table}[t!]
\centering
\captionof{table}{ Circuit parameters in equivalent circuit model of experimental 5-coil linear array}
\vspace{2mm}
\setlength{\tabcolsep}{4pt}
\begin{tabular}{l c c c c c}
     \hline
     \noalign{\smallskip}
     \textbf{Circuit} & $1$ & $2$ & $3$ & $4$ & $5$ \\
     \noalign{\smallskip}
    \hline
    \noalign{\smallskip}
    $L\ (\mu H)$ & $16.7$ & $17.4$ & $12.7$ & $17.3$ & $17.9$ \\
    \noalign{\smallskip}
    $C\ (nF)$ & $1.72$ & $1.72$ & $1.73$ & $1.72$ & $1.72$ \\
    \noalign{\smallskip}
    $R\ (\Omega)$ & $3.73$ & $4.0$ & $3.79$ & $3.88$ & $3.96$ \\
    \noalign{\smallskip}
    
    \hline
    \end{tabular}
    \label{tab:5coilvals}
\end{table}

Completely identical inductors in an array theoretically result in the central coil having the highest effective inductance (as indicated by the steeper gradients in impedance at low frequencies for the middle coils), seen in the theoretical three and five coil arrays (Figures~\ref{fig:exampleZf} and \ref{fig:5coil_preds} respectively). Meanwhile, dissimilar array elements seem to result in the central coil exhibiting the lowest effective inductance. The mechanism for this is not clear.

There is an additional effect which has not been accounted for in the theoretical circuit models and which is likely to be contributing to the disparity in impedance spectra between the experiment and the circuit theory when the coils are identical. That is, that placing ferrite cored coils next to each other will impact on the inductance of each array element. 
If the manually modified values of the circuit modelled asymmetric array are similar to the actual values, then the pattern of inductances within the linear array suggests that the coils at the end have a smaller increase in inductance when placed in the array, compared to coils 2 and 4.  This makes physical sense given that coils 2 and 4 have ferrite cores to both sides of them, while coils 1 and 5 have ferrite coils only on one side.  Future studies will explore how this can be mitigated through the use of dummy cores at the ends of the array and through close-packing these arrays, and parameter optimisation used to match theory to experimental results.

Figure~\ref{fig:five-coil-exp}.c shows the experimentally measured voltages across the array elements at each resonant frequency (i) compared to the eigenmodes predicted by the asymmetric array circuit model (ii).  
In spite of the asymmetric array configuration, the modes remain recognisable to those shown for the idealised identical element system shown in Figure~\ref{fig:5coil_preds}.c, but exhibit some distorted features.  
The experimental and approximated circuit model results exhibit highly comparable modes at their respective resonant frequencies, with a few minor exceptions occurring for coils 1 and 2 in mode $f_2$, and coils 4 and 5 in mode $f_4$.  This is likely in part related to the close proximity of these resonant frequencies to other resonant modes.

\section{Conclusion} \label{sec:conclusion}
This paper presents and validates a circuit theory for predicting and evaluating the resonant phenomena observed in over-coupled resonant inductive arrays. The mathematical formalisation is presented for electrical measurements across a single parallel resonant coil element in an array of passive series resonant inductors, and critically evaluated against finite element simulations and experimental results of linear resonant inductor arrays.

The circuit model, FE simulations and experimental results all confirm the unique multi-modal resonant phenomena observed in measurements of individual array elements, with these resonant frequencies corresponding to specific magnetic excitation "modes"; equivalent to vibrational modes in multi-degree-of-freedom systems. 

With the models presented, novel inductive arrays can be designed and optimised to exploit the unique resonant effects of over-coupling for applications in wireless power transfer and array sensing.  
The ability to identify and predict resonant modes in arrays opens the door for array sensing without electronic multiplexing of the whole array, but rather through the excitation of individual coils within the array by excitation at specific frequencies.  
This interpretation of magnetic modes has the potential to enable magnetic field shaping through simple frequency selection.

Further work is required to explore the many and varied possible spatial array configurations and to apply these resonant array principles to novel applications.  It is anticipated that the theory and analysis presented herein has the potential to initiate a unique field of research and a profound change in how arrayed inductor systems are designed and used.

\bibliographystyle{plain}
\bibliography{library}

\appendix

\end{document}